\documentclass[epsfig,aps,onecolumn,superscriptaddress,footinbib,preprint,amsmath,amssymb]{revtex4}

\usepackage{graphicx}% Include figure files

\begin{document}
\title{Time-resolved resonance and linewidth of an ultrafast switched GaAs/AlAs microcavity}
\author{Philip J. Harding}
\affiliation{$^1$ Center for Nanophotonics, FOM Institute for Atomic and Molecular Physics (AMOLF),
Kruislaan 407, 1098 SJ Amsterdam, The Netherlands}%
\affiliation{$^2$Complex Photonic Systems (COPS), MESA+ Institute for Nanotechnology, University of Twente, 7500AE Enschede, The Netherlands}
\email{p.j.harding@alumnus.utwente.nl}
\author{Allard P. Mosk}
\affiliation{$^2$Complex Photonic Systems (COPS), MESA+ Institute for Nanotechnology, University of Twente, 7500AE Enschede, The Netherlands}
\author{Alex Hartsuiker}
\affiliation{$^1$ Center for Nanophotonics, FOM Institute for Atomic and Molecular Physics (AMOLF),
Kruislaan 407, 1098 SJ Amsterdam, The Netherlands}%
\author{Yoanna-Reine Nowicki-Bringuier}
\affiliation{$^3$ CEA/INAC/SP2M, CEA-CNRS Nanophysics and Semiconductor Laboratory, 17 rue des Martyrs, 38054 Grenoble Cedex, France}
\author{Jean-Michel G\'erard}
\affiliation{$^3$ CEA/INAC/SP2M, CEA-CNRS Nanophysics and Semiconductor Laboratory, 17 rue des Martyrs, 38054 Grenoble Cedex, France}
\author{Willem L. Vos}
\affiliation{$^1$ Center for Nanophotonics, FOM Institute for Atomic and Molecular Physics (AMOLF),
Kruislaan 407, 1098 SJ Amsterdam, The Netherlands}%
\affiliation{$^2$Complex Photonic Systems (COPS), MESA+ Institute for Nanotechnology, University of Twente, 7500AE Enschede, The Netherlands}

\begin{abstract}
We explore a planar GaAs/AlAs photonic microcavity using pump-probe spectroscopy. Free carriers are excited in the GaAs with short pump pulses.
The time-resolved reflectivity is spectrally resolved short probe pulses. We show experimentally that the cavity resonance and its width depend
on the dynamic refractive index of both the $\lambda$-slab and the $\lambda/4$ GaAs mirrors. We clearly observe a double exponential relaxation
of both the the cavity resonance and its width, which is due to the different recombination timescales in the $\lambda$-slab and the mirrors. In
particular, the relaxation time due to the GaAs mirrors approaches the photon storage time of the cavity, a regime for which nonlinear effects
have been predicted. The strongly non-single exponential behavior of the resonance and the width is in excellent agreement to a transfer-matrix
model taking into account two recombination times. The change in width leads to a change in reflectivity modulation depth. The model predicts an
optimal cavity $Q$ for any given induced carrier density, if the modulation depth is to be maximized.
\end{abstract}

\maketitle

\section{Introduction}
Recently, there has been a fast-growing interest to switch cavities on ultrafast timescales. Switching cavities is important for dynamical
control of light, such as optical wavelength modulation, bandwidth conversion, density of states switching, and trapping and releasing photons
[1-5]. Micro- and nanocavities are particularly attractive in this context, as they are
amenable to be integrated in next generation all-optical networks. Cavity switching becomes especially interesting when the photon storage
time $\tau_{\rm{ph}}$ becomes comparable to the timescales on which the cavity resonance changes \cite{Notomi:06}. In this regime, strong pulse
chirping and frequency conversion effects have been observed \cite{Yacomotti:06, Preble:07}.

The availability of strong, short pulses, the good reproducibility, and the possibility of integration in all-optical networks has furthered the
popularity of free-carrier switching of resonators or cavities, see, e.g., Refs. \cite{Almeida:04, Harding:07, Foerst:07}. The dispersion of
these carriers result in a change of both real ($n'$) and imaginary ($n''$) part of the refractive index that is proportional to the induced
carrier density $N$. The carriers recombine within a typical timescale ranging from ps to ns, and the cavity resonance relaxes to its unswitched
value. Surprisingly, these relaxation and broadening mechanisms have hardly been discussed, mostly due to the lack of frequency- and
time-resolved data. While other groups have limited their studies to either the reflectivity at two frequencies [3, 11-14] or to the reflectivity at two probe delays \cite{Fushman:07}, we present here a systematic study at {\em{all}}
frequencies and delays which allows us to investigate the dynamic resonance and lineshape. In our earlier contribution, insufficient spectral
resolution had prevented us from investigating these broadening mechanisms \cite{Harding:07}. Here, we study both the dynamic shifting and
broadening of a cavity resonance by performing time- and frequency resolved pump-probe spectroscopy on a planar GaAs/AlAs microcavity, with a
frequency resolution much higher than the dynamic cavity linewidth.

The most general cavity is a slab of a dielectric, bounded by two mirrors, see Fig. \ref{fig:Schematic}(a). This well-known Fabry-P\'erot cavity serves as the basis for the 
understanding of 1-, 2- and 3D cavities  \cite{Vahala:03}. Many important physical properties, such as storage time, bandwidth, mode volume and
reflectivity depend on the optical properties of both the slab and the mirrors, and can be calculated analytically for such a model system
\cite{Born:97}. Therefore, to obtain a physical understanding of switching of advanced photonic crystal micro- and nanocavities, it is
advantageous to consider first the Fabry-P\'erot cavity. One of such processes is switching by exciting free carriers in the mirrors and the
$\lambda$-slab with a strong pump pulse, whereby the cavity resonance changes rapidly within $\tau_{\rm{on}}$, and relaxes back to its original
state within $\tau_{\rm{off}}$.  Previously, nonlinear effects observed by Refs. \cite{Yacomotti:06, Preble:07} were obtained during the
notoriously uncontrollable rapid up-change time of the cavity resonance, as $\tau_{\rm{on}}$ depends on material parameters. Therefore, many
efforts are devoted to decreasing the timescale of $\tau_{\rm{off}}$ \cite{Foerst:07, Tanabe:07, Chin:96, Segschneider:97}. Here, we show how
the relaxation time can quite easily approach the storage time of the cavity, and show how this can be explained in terms of the basic
Fabry-P\'erot cavity. Nonlinear effects for optimized planar microcavity samples are expected during $\tau_{\rm{off}}$. Switching of planar
microcavities is thus expected to be a step towards achieving nonlinear pulse phenomena.

\section{Experimental setup}
\begin{figure}
\begin{center}
\includegraphics[width=9cm]{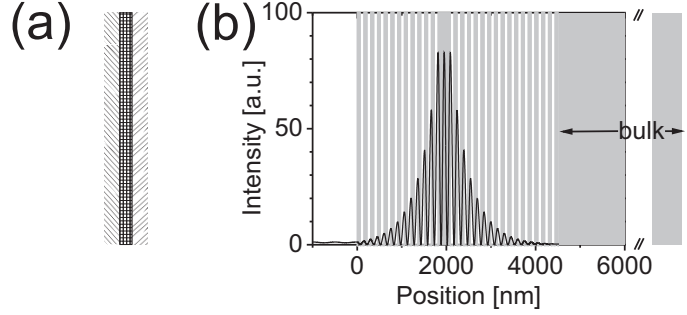}
\caption{(a) The model Fabry-P\'erot system, consisting of a slab bounded by two mirrors. The resonance depends on both the optical properties of the mirrors and of the slab. (b) Schematic of sample. The GaAs is indicated by the shaded areas. The GaAs $\lambda$-slab is bounded by 12 and 16 pairs of $\lambda/4$ GaAs/AlAs pairs at the front and at the rear side, respectively. The intensity distribution is shown at the cavity resonance, for a unity field.}
 \label{fig:Schematic}
 \end{center}
\end{figure}

Our sample is a planar microcavity consisting of a GaAs $\lambda$-slab sandwiched between two Bragg mirrors, which was also studied earlier
\cite{Harding:07}. The sample is schematically shown in Fig. \ref{fig:Schematic}(b). The slab has a thickness of $275.1 \pm 0.1$ nm, and the Bragg mirrors consist of 12 and 16 pairs of $\lambda/4$ thick layers
of nominally pure GaAs or AlAs. The sample gives rise to a cavity resonance at $E_{\rm{cav}} = 1.2783$ eV, well below the bandgap of GaAs ($E_g = 1.44$
eV) to avoid intrinsic absorption at the cavity resonance. The sample is grown with molecular beam epitaxy at 550$^{\circ}$C to optimize the optical quality, but no effort was put into reducing the recombination time. A slight
variation of a few $\%$ in stopgap and cavity resonance over the sample allowed us to verify the experimental observations for different
resonance frequencies. We performed two separate studies which yielded consistent results. For experiments outside the present scope the sample
was doped with $10^{10}$cm$^{-2}$ InGaAs/GaAs quantum dots, which hardly influence our experiment \cite{InfluenceQD}. The experiments were conducted at room temperature. Continuous-wave
(cw) reflectivity was measured with a broad band white light setup with a spectral resolution of $\sim 0.25$ meV.

Our switching setup consists of two optical parametric amplifiers (OPA, Topas), that are the sources of the pump and probe beams
\cite{EuserWoodpile:08}. The frequency of both OPAs are computer controlled and have a continuously tunable output frequency between 0.44 and
2.4 eV. The repetition rate is $f_{\rm{rep}} = 1$ kHz. In order to verify that the cavity was not degrading due to heating effects, we
simultaneously measured both switched and unswitched spectra by using a chopper. This averaging caused the effective pump or probe rate to be
$f_{\rm{rep}}/2 = 500$ Hz. The alignment and synchronization of the pulses to the chopper was chosen so as to overlap only every second pump
pulse with a probe pulse, so that the effective pump rate was $f_{\rm{rep}}/4$ \cite{EuserThesis:07}. The pulse duration is $\tau_P = 140 \pm
10$ fs (measured at $E_{\rm{Pump}} = 0.95$ eV) \cite{MeasurementtauP}, and the spectral width is $\Delta E/E_0 = 1.33 \%$. A 40 cm long delay stage controls the probe delay with a time resolution of
$\Delta t = 10$ fs. The pump beam has a much larger Gaussian focus of 113 $\mu$m FWHM than the probe beam (28 $\mu$m), ensuring that only the
central flat part of the pump focus is probed. The probe fluence of $I_{\rm{Probe}} = 1 \pm 0.3$ mJcm$^{-2}$ is kept well below the pump fluence
of $I_{\rm{Pump}} = 30 \pm 3$ mJcm$^{-2}$ to prevent inadvertent pumping by the probe pulses. These high pump fluences, which shift the cavity
resonance by several linewidths, facilitate the analysis of the dynamic behavior. Free carriers are excited in the GaAs by two-photon absorption at $E_{\rm{pump}} = 0.73$ eV to obtain a spatially homogeneous distribution of carriers \cite{Euser:05}.

In order to resolve the dynamic lineshape of the microcavity, the reflected probe light was analyzed with a spectrometer. This spectrometer
consists of a PI/Acton SP-2558 spectrograph, using a 1024 channel InGaAs detector (OMA-V), yielding a resolution of $0.12$ meV at $1.24$ eV,
much higher than the unswitched cavity linewidth of $\Delta_0 = 1.03$ meV. To average over pulse-to-pulse variations of the OPAs, the
spectrograph was operated in free-running mode for a duration of $1$ s, acquiring 500 probe pulses, of which which half of them are incident
together with a pump pulse.

\section{Linear reflectivity}
\begin{figure}
\begin{center}
\includegraphics[width=7cm]{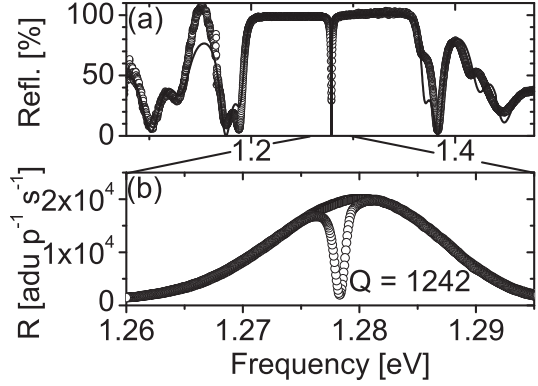}
\caption{(a) Continuous wave reflectivity spectrum of the planar microcavity (open circles). The cavity resonance can clearly
be seen at $E_{\rm{cav}} = 1.279$ eV. The solid curve is a transfer matrix calculation that includes the dispersion and absorption of GaAs and
AlAs. (b) Reflectance spectrum in analog to digital units (adu) per pixel per second from the same microcavity (circles) measured with the
pulsed probe beam ($\tau_P = 140 \pm 10$ fs, $\Delta \omega_P = 1.35 \%$), and from a reference (squares).}
 \label{fig:ProbeRefl}
 \end{center}
\end{figure}

Fig. \ref{fig:ProbeRefl}(a) shows a white light reflectivity spectrum of the planar photonic microcavity at normal incidence. The high peak
between 1.193 and 1.376 eV is due to the stopgap of the Bragg stacks. The stopband has a broad width of 0.183 eV ($14.3 \%$ relative bandwidth),
which confirms the high photonic strength. Near $1.279$ eV we observe a sharp resonance caused by the $\lambda$-slab in the structure. This
resonance has a width of $1.7$ meV that is limited by wavevector spreading caused by the finite numerical aperture of 0.12. A transfer matrix
(TM) calculation was performed for the reflectivity. In the calculation, we included the dispersion and absorption of GaAs \cite{Blakemore:82}
and AlAs \cite{Fern:71}. The only free parameters in the model were the thicknesses of the GaAs ($d_{\rm{GaAs}} = 68.78 \pm 0.03$ nm) and AlAs
($d_{\rm{AlAs}} = 81.90 \pm 0.03$ nm). The results reproduce the experimental resonance, stopband, and Fabry-P\'{e}rot fringes well.

We made use of the large probe bandwidth compared to the cavity linewidth to spectrally resolve the microcavity with the spectrometer (Fig.
\ref{fig:ProbeRefl}(b)): The spectral width of the cavity at the resonance $E_{\rm{cav}} = 1.2783$ eV is $\Delta_0 = 1.03$ meV, and corresponds
to an unswitched quality factor of $Q_0 = 1242$ and thus a photon storage time of $\tau_{\rm{ph}} = 640$ fs. This larger quality factor with
respect to the cw measurement is attributed to the reduced wavevector spreading due to the lower numerical aperture (NA = 0.02) which is
confirmed by the higher modulation depth. The large probe bandwidth in combination with the high spectral resolution allows us to probe not only
the cavity resonance as in \cite{Harding:07} but also the lineshape, even when the cavity resonance has shifted by several linewidths.

\section{Results and discussions}
\begin{figure}
\begin{center}
\includegraphics[width=7cm]{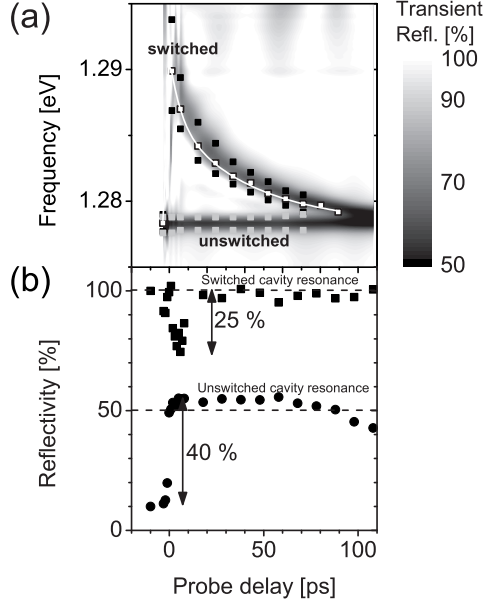}
\caption{(a) Time- and frequency resolved reflectivity spectra of both switched and unswitched cavity vs. probe delay, measured alternatingly.
The white squares indicate the frequency of the switched cavity resonances, the white curve is a guide to the eye. The half-minima of the
unswitched cavity resonance are given by the grey squares. The half-minima of the switched cavity resonance are given by the black squares, and
the pump and probe fluences are $I_{\rm{Pump}} = 30 \pm 3$ mJcm$^{-2}$ and $I_{\rm{Probe}} = 1 \pm 0.3$ mJcm$^{-2}$. (b) Ultrafast change in
reflectivity at the unswitched (1.2783 eV) and at the switched (1.2898 eV) cavity resonance. The observed changes should be multiplied by a
factor of 2 to correct for averaging the switched and unswitched spectra.} \label{fig:TransientRefl}
\end{center}
\end{figure}

We dynamically probe the excited cavity by time- and frequency resolved pump-probe reflectivity. Figure \ref{fig:TransientRefl}(a) shows the
transient reflectivity spectrum as a function of probe delay. The unswitched cavity resonance is clearly seen at 1.2783 eV. The width of the
unswitched cavity resonance does not increase in the slightest, as shown by the half maxima. We therefore conclude that there are no permanent
damage or heating effects to the sample.

In Fig. \ref{fig:TransientRefl}(a), the second trough emerging at 1.2898 eV, at $\Delta t = 3$ ps, is the switched cavity resonance. Up to probe
delays $\Delta t = \tau_{\rm{on}} = 3$ ps, the cavity resonance shifts by as much as $\Delta E_{\rm{cav}} = 11.5$ meV, or 11 cold cavity
linewidths. This switching-on time is in good agreement to the thermalization time of free carriers in GaAs \cite{Hense:97, Cho:90}. Both the
resonance energy and linewidth relax to their initial values within $\Delta t = 100$ ps. Both the switched resonance's half maxima as well as
their resonances are indicated. We observe that the dynamic resonance is not centered between the half maxima, but is located much closer to the
red edge. The asymmetric shape of the line indicates that resonance is shifting appreciably during the dynamic photon storage time, an effect
which we coin 'kinetic broadening'. Therefore, the broad width of $\Delta = 6$ meV at $\tau_{\rm{on}}$ is partially caused by this artificial
kinetic effect. The real width is expected to be significantly lower.

We plot the transient reflectivity at both the unswitched and the switched cavity resonance in Fig. \ref{fig:TransientRefl}(b). At the
unswitched cavity resonance, the reflectivity increases from $10 \%$ at $\Delta t = -3$ ps to just over $50 \%$ at $\Delta t = 3$ ps. This
corresponds to a modulation depth of $2(50 \% - 10 \%) = 80 \%$, where the factor 2 corrects for averaging over switched and unswitched spectra.
At the switched cavity resonance however, the modulation depth is $2(100\% - 75\%) = 50 \%$. We can thus see that broadening and the concomitant
decrease in reflectivity leads to a significantly reduced signal modulation depth.

Before we quantitatively describe the behavior of the modulation depth, we will focus on the dynamics of the cavity resonance. Figure
\ref{fig:RelAndAbsResAndWidth}(a) shows the time-resolved frequency shift of the cavity resonance, taken from Fig. \ref{fig:TransientRefl}(a).
Naively, one might assume that the behavior of the cavity resonance follows a single exponential decay \cite{Almeida:04,
Harding:07, Tanabe:07, Fushman:07}. Interestingly, Fig. \ref{fig:RelAndAbsResAndWidth}(a) shows a fit of a TM model to the resonances assuming one single recombination time
for both the $\lambda$-slab and the GaAs mirrors, but the agreement is only moderate. Permitting the recombination times in the $\lambda$-slab to differ from those
in the GaAs mirrors, the fit is much better (details below): Because the resonant mode extends into the Bragg mirrors for a distance $L_B = 425$
nm, the cavity resonance and the width will be a function of the dynamic refractive index in both the $\lambda$-slab and the Bragg mirrors. At
pump incidence, the carrier density and thus the refractive index will be the same everywhere. At probe delays $\Delta t > \tau_{\rm{on}}$,
however, the carriers in the thin $\lambda/4$ Bragg recombine more rapidly than in the thicker $\lambda$ slab. Therefore, the time dependent
refractive indices will differ, and will cause the resonance and width to behave as a double exponential. Returning to the analogy with the
Fabry-P\'erot model, we identify the $\lambda$-slab of the microcavity with the slab, while the Bragg mirrors are identified with the two
mirrors. Because the Fabry-P\'erot resonator is the template for {\em{all}} cavities, we conclude that our observation of two separate relaxing sub-systems will be valid for {\em{all}} micro- and nanocavities which consist of topologically separated slab and mirrors. 

\section{Model and interpretation}
\begin{figure}
\begin{center}
\includegraphics[width=7cm]{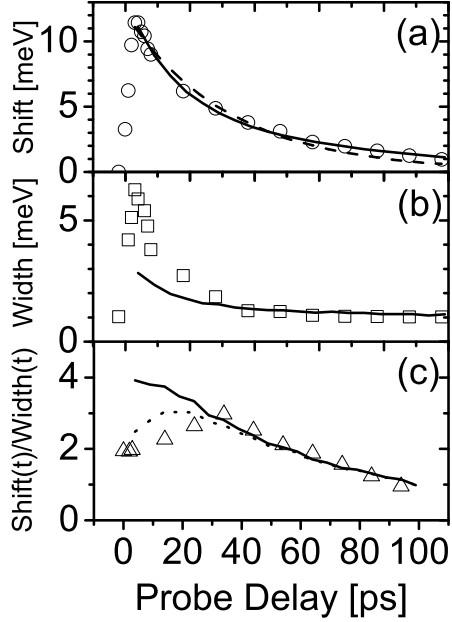}
\caption{All data extracted from Fig. \ref{fig:TransientRefl}. (a) Time-resolved shift of the cavity resonance (open circles). The expected
resonance from a TM model assuming equal recombination times in both the GaAs mirrors and the $\lambda$-slab $\tau_{\rm{gc}} = \tau_{\rm{gm}} = 38 \pm 0.5$ ps is given by the dashed curve. The solid curve is a TM calculation for $N_0 = 2.58 \times 10^{19}$ cm$^{-3}$,
$\tau_{\rm{gm}} = 14.7 \pm 0.5$ ps and $\tau_{\rm{cav}} = 62.9 \pm 1$ ps. (b)
Time-resolved width of the cavity (open squares). Here, the solid curve is a TM calculation including free-carrier absorption from the TM calculation, spatial
inhomogeneity, wavevector spreading and kinetic broadening, which includes a time-dependent storage time $\tau_{\rm{ph}}$ as shown in eq. \ref{eq:kin}. (c) Open triangles: Measured shift of cavity resonance ratioed by the measured width. Solid curve: Calculated dynamic shift divided by dynamic width (open triangles) from (a) and (b), i.e., assuming a variable storage time $\tau_{\rm{ph}}$. The dotted curve is a calculation of the shift-width ratio, where the storage time is constant in the calculation of the width. } \label{fig:RelAndAbsResAndWidth}
\end{center}
\end{figure}

We will now quantitatively describe the relaxation of the cavity resonance. To a first order approximation, the change in dynamic cavity
resonance is a linear function of the refractive index in the GaAs mirrors $n_{\rm{gm}}$ and in the GaAs $\lambda$-slab $n_{\rm{gc}}$, which are
weighted by field distribution coefficients $a$ and $b$, respectively \cite{LinearizationN}. If the field distribution does not change significantly during the switch \cite{ChangeLB}, we can write for the shift in cavity resonance:
\begin{equation}
\Delta E_{\rm{cav}}(\Delta t) = a \Delta n'_{\rm{gm}}(\Delta t) + b \Delta n'_{\rm{gc}}(\Delta t).
\label{eq:linearEcav}
\end{equation}
Since the induced change in refractive index is proportional to the carrier density $N_0$ for sufficiently small $N_0$, equation
\ref{eq:linearEcav} can be written as
\begin{equation}
\Delta E_{\rm{cav}}(\Delta t) = a' N_{\rm{gm}}(\Delta t) + b' N_{\rm{gc}}(\Delta t),
\label{eq:EcavN}
\end{equation}
where $a'$ and $b'$ are field weighting factors in the mirrors and in the $\lambda$-slab, respectively. Here, $a' = \frac{a''e^2}{2m_e^* n_g \epsilon_0 \omega_{\rm{pr}}^2}$ and similarly for $b'$, where $m_e^*$ is the
effective electron-hole mass in GaAs, $n_{\rm{g}}$ the static refractive index of GaAs, $\epsilon_0$ the static dielectric constant of GaAs, and
$\omega_{\rm{pr}}$ the probe frequency in rad/s. Often, the time evolution of $N_{\rm{gm}}(t)$ and $N_{\rm{gc}}(t)$ are considered to be equal.
Due to the different thicknesses, however, they are governed by different relaxation constants $\tau_1$ and $\tau_2$. Therefore, the change in
cavity resonance can be written as
\begin{equation}
 \Delta E_{\rm{cav}}(\Delta t) = a' N_0\exp(-\Delta t/\tau_1) + b' N_0\exp(-\Delta t/\tau_2),
  \label{eq:DoubleDecay}
\end{equation}
Because of homogeneous carrier excitation via two-photon absorption \cite{Euser:05}, the carrier density is initially constant. In the present experiment, non radiative recombination of free carriers at GaAs/GaAlAs interfaces is expected to be the dominant recombination process
\cite{Gerard:90}. This enhanced recombination which takes place at the planar surfaces, combined with carrier diffusion, and topologically separated $\lambda$-slab and mirrors resulting in a discontinuous carrier density distribution, gives rise to the double exponential behavior. The different rates are thus identified with the different recombination rates in
the GaAs mirrors and in the $\lambda$-slab, due to the different thicknesses. From the fit, using 100 iterations with the Levenberg-Marquardt algorithm, and assuming $N_0 = 2.58 \times 10^{25}$m$^{-3}$ (see below), we find $\tau_1 = 5.5 \pm 1$ ps and $\tau_2 = 44.8
\pm 2$ ps, while $a' = 1.3 \pm 0.1 \times 10^{-28}$ and $b' = 3.2 \pm 0.1 \times 10^{-28}$ eVm$^3$. From the relative sizes of $a'$ and $b'$ we observe that the field is primarily located in the $\lambda$-slab. To calculate the {\em{recombination}} times instead of the
relaxation times, which also takes into account the small change in field distribution, it is necessary to fit a series of TM spectra to our
measured data. Here, we fit the initial carrier density $N_0$, and two recombination times $\tau_{\rm{gm}}$ and $\tau_{\rm{gc}}$ for the
recombination times in the GaAs mirrors and the $\lambda$-slab, respectively. The fit gives $N_0 = 2.58 \pm 0.01 \times 10^{19}$ cm$^{-3}$,
$\tau_{\rm{gm}} = 14.8 \pm 0.5$ ps and $\tau_{\rm{cav}} = 62.9 \pm 1$ ps, and is shown in Fig. \ref{fig:RelAndAbsResAndWidth}(a). The sum of the differences squared is 44.6, in contrast to 166.1 if we assumed that the dynamic carrier density is the same in the mirrors and the
$\lambda$-slab ($\tau_{\rm{gc}} = \tau_{\rm{gm}} = 38 \pm 0.5$ ps, dashed curve). Comparing $\tau_{\rm{gm}}$ and $\tau_{\rm{gc}}$ to $\tau_1$ and
$\tau_2$, respectively, we find that the former are considerably higher, which we attribute to the change in field
distribution. Note that the ratio of fitted recombination times $\tau_{\rm{cav}}/\tau_{\rm{gm}}$ is in excellent agreement with the ratio of
layer thicknesses $\lambda/(\lambda/4)$, confirming the notion of the double exponential decay, due to the dominant recombination at the planar GaAs/AlAs
interfaces.

We emphasize that these two different recombination rates can significantly change the analysis of the dynamic field behavior in the cavity:
even though an effective single exponential decay time can be orders of magnitude above the cavity dwell time, the composite shorter
recombination times will give rise to significantly different field dynamics \cite{Notomi:06, Yacomotti:06}. To gain access to timescales
comparable to $\tau_{\rm{ph}}$, several groups made use of the fast upswitch time $\tau_{\rm{on}} \sim 1$ ps \cite{Notomi:06, Preble:07}. Here,
we probe for the first time during the relaxation, whose timescale is controllable, and still comparable to $\tau_{\rm{ph}}$. The recombination
rates in microcavities may further be increased controllably by growing samples with a larger number of recombination centers at the GaAs/AlAs
interfaces \cite{Segschneider:97}. Interestingly, Tanabe {\em{et al.}} deduced a non-single exponential cavity resonance relaxation of a 2-D hexapole cavity by rigorously solving the diffusion equation combined with surface recombination \cite{Tanabe:08}. For their system, in which slab and the mirrors are topologically connected, the different relaxation rates are related to the carrier diffusion and the surface recombination rates, respectively. 

In order to quantify the width (Fig. \ref{fig:RelAndAbsResAndWidth}(b)), which is crucial for understanding the behavior of the modulation
depth, we calculate the dynamic width from the TM model using the carrier densities obtained previously. In addition to the time-dependent
broadening by the free carriers, we include three additional broadening mechanisms inherent to our experiment: first, wavevector spreading
because of the finite probe NA, and second, the spatially changing resonance at the focus waist. These two broadening mechanisms are static, and
result in the calculated width approximately matching the measured width. Finally, we take into account kinetic broadening: the spectrometer
measures all light exiting the shifting cavity. The total width including kinetic broadening is given by
\begin{eqnarray}
\Delta_{\rm{kin}}(t) &=& \left(\Delta_{\rm{inhom}}(t)^2 + \left[\tau_{\rm{ph}}(t)\frac{d}{dt}E_{\rm{cav}}(t)\right]^2\right)^{1/2} \\
&=& \left(\Delta_{\rm{inhom}}(t)^2 + \left[\hbar N_0/(e \Delta_0(t))\left(a'/\tau_1\exp(-\Delta t/\tau_1) + c'/\tau_2\exp(-\Delta t/\tau_2)\right)\right]^2\right)^{1/2}
\label{eq:kin}
\end{eqnarray}
Using $\tau_{\rm{gm}}$ and $\tau_{\rm{gc}}$ obtained from the fit in Fig. \ref{fig:RelAndAbsResAndWidth}(a), and the relaxation parameters, we
can calculate the dynamic width (Fig. \ref{fig:RelAndAbsResAndWidth}(b), solid line). At $\Delta t = 3$ ps, the calculated width is only around
half of the measured width. At larger probe delays $\Delta t > 20$ ps, the calculation agrees excellently with the measurement. The disagreement
at short probe delays might stem from the fact that the storage time $\tau_{\rm{ph}}(t)$ stays constant throughout the excitation and the subsequent relaxation. In that case, the kinetic
contribution might be considerably larger. To elaborate this point, we plot the dynamic shift ratioed by the dynamic width, shown in Fig. \ref{fig:RelAndAbsResAndWidth}(c). At coincidence, the ratio is 2. It shows a maximum at $\Delta t = 35$ ps, before relaxing to $0$ at longer probe delays. The agreement of the calculation to the data is excellent for $\Delta t > 20$ ps, but fails at short probe delays. We can exclude coherent transients and electron-electron thermalization due to the short times (in the order of $\sim 3$ ps \cite{Cho:90} and $100$ fs \cite{Sabbah:02}, respectively). Transient heating can be ruled out because heating would also influence the cavity resonance, which we find to agree excellently to our model without heating. However, assuming a {\em{constant}} $\tau_{\rm{ph}}(t)$ (Fig. \ref{fig:RelAndAbsResAndWidth}(c), dotted line), the agreement is excellent for both $\Delta t > 20$ ps and $\Delta t < 20$ ps, indicating that $\tau_{\rm{ph}}$ might indeed remain constant. We note that this surprising finding should be explored further, as this opens the possibility of switching with only limited degradation of the storage time.

\begin{figure}
\begin{center}
\includegraphics[width=7cm]{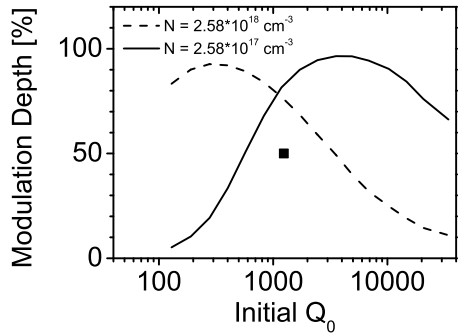}
\caption{Semi-log plot of calculated modulation depth vs. different initial $Q_0$ at the switched cavity resonance for a constant carrier
density. Square: our measurement.}
\label{fig:ModDepth}
\end{center}
\end{figure}

In order to show by how much the modulation depth is affected by free carrier broadening, we plot in Fig. \ref{fig:ModDepth} the expected
modulation depth at the switched cavity resonance at pump-probe coincidence for different initial $Q_0$. Here, the carrier density is kept
constant at $N_0 = 2.58 \times 10^{19}$ cm$^{-3}$, as in the preceding experiment. At low values of $Q_0$, the modulation depth is shallow, as
expected. It acquires a maximum at around $Q_0 = 300$. Up to this point, the modulation depth has been limited by the low $Q_0$ of the sample.
For $Q_0 > 300$, the expected modulation depth decreases rapidly. In this regime, the linewidth of the resonance becomes more and more sensitive
to free carrier absorption, and to the decrease of the mirrors' reflectivity. The measured modulation depth of the present measurement is
included. The discrepancy indicates that the modulation depth is not well understood for short probe delays. For a carrier density an order of
magnitude lower, the maximum is reached for $Q_0 = 5000$. For $Q_0 > 5000$, the sensitivity of the modulation depth on $Q_0$ is less pronounced.
From the calculations we conclude that there is a maximum attainable modulation depth for every initial linewidth and carrier density.

\section{Conclusion}
Using time- and frequency resolved pump-probe spectroscopy, we have experimentally demonstrated ultrafast switching of up to 11 linewidths
within 3 ps of a microcavity using two-photon absorption. The dynamics of the cavity resonance is investigated in detail, where we find a double
exponential behavior, which is shown to result from the different recombination rates in the $\lambda$-slab and the Bragg mirrors. We suggest
that {\em{all}} photonic crystal cavities are subject to at least two different relaxation times, a fact which has not been measured
previously. Our observed distribution of recombination times is predicted to strongly modify the field dynamics with respect to a single-exponential
decay. In particular, the regime for which the timescale of relaxation of the cavity resonance is comparable to the photon dwell time becomes
accessible experimentally. From the behavior of the resonance, the time-resolved width can be deduced. We discuss broadening effects due to wavevector spreading, spatially varying resonances on the sample, and kinetic broadening. Surprisingly, the calculations, which assume a constant storage time $\tau_{\rm{ph}}$ during the excitation and subsequent relaxation of the cavity resonance agree best to the data, a fact which has not been appreciated previously. The broadening mechanisms are shown to contribute to the final width at all probe delays. Finally, we discuss the effect of these broadening mechanisms on the modulation depth. From a calculation we are able to
infer an optimum value of $Q_0$ for a given carrier density and a required modulation depth.

\section*{Acknowledgements}
We thank Bart Husken for help with the spectrometer, and the two reviewers for their comments. This work is part of the research programme of the "Stichting voor Fundamenteel Onderzoek
der Materie" (FOM), which is financially supported by the NWO. This research was also supported by NanoNed, a nanotechnology programme of the
Dutch Ministry of Economic Affairs, and by a VICI fellowship from the "Nederlandse Organisatie voor Wetenschappelijk Onderzoek" (NWO) to WLV.


\begin{thebibliography}{10}
\newcommand{\enquote}[1]{``#1''}

\bibitem{Johnson:02}
P.~M. Johnson, A.~F. Koenderink, and W.~L. Vos, \enquote{Ultrafast switching of
  photonic density of states in photonic crystals,} \prb. \textbf{66},
  081102(R) (2002).

\bibitem{Bret:03}
B.~P.~J. Bret, T.~L. Sonnemans, and T.~W. Hijmans, \enquote{Capturing a light
  pulse in a short high-finesse cavity,} \pra \textbf{68}, 023807
  (2003).

\bibitem{Almeida:04}
V.~R. Almeida, C.~A. Barrios, R.~R. Panepucci, and M.~Lipson,
  \enquote{All-optical control of light on a silicon chip,} \nat
  \textbf{431}, 1081 -- 1084 (2004).

\bibitem{Xu:07}
Q.~Xu, P.~Dong, and M.~Lipson, \enquote{Breaking the delay-bandwidth limit in a
  photonic structure,} Nat. Phys. \textbf{31}, 406--410 (2007).

\bibitem{Tanaka:07}
Y.~Tanaka, J.~Upham, T.~Nagashima, T.~Sugiya, T.~Asano, and S.~Noda,
  \enquote{Dynamic control of the q factor in a photonic crystal nanocavity,}
  Nat. Mater. \textbf{6}, 862--865 (2007).

\bibitem{Notomi:06}
M.~Notomi and S.~Mitsugi, \enquote{Wavelength conversion via dynamic refractive
  index tuning of a cavity,} \pra \textbf{73}, 051803 (2006).

\bibitem{Yacomotti:06}
A.~M. Yacomotti, F.~Raineri, C.~Cojocaru, P.~Monnier, J.~Levenson, and R.~Raj,
  \enquote{Nonadiabatic dynamics of the electromagnetic field and charge
  carriers in high-{Q} photonic crystal resonators,} \prl
  \textbf{96}, 093901 (2006).

\bibitem{Preble:07}
S.~F. Preble, Q.~Xu, and M.~Lipson, \enquote{Changing the colour of light in a
  silicon resonator,} Nat. Phot. \textbf{1}, 293--296 (2007).

\bibitem{Harding:07}
P.~J. Harding, T.~G. Euser, Y.~R. Nowicki-Bringuier, J.-M. G\'erard, and W.~L.
  Vos, \enquote{Ultrafast optical switching of planar {GaAs/AlAs} photonic
  microcavities,} \apl \textbf{91}, 111103 (2007).

\bibitem{Foerst:07}
M.~F\"orst, J.~Niehusmann, T.~Pl\"otzing, J.~Bolten, T.~Wahlbrink, C.~Moormann,
  and H.~Kurz, \enquote{High-speed all-optical switching in ion-implanted
  silicon-on-insulator microring resonators,} \ol \textbf{32},
  2046--2048 (2007).

\bibitem{Tanabe:07}
T.~Tanabe, K.~Nishiguchi, A.~Shinya, E.~Kuramochi, H.~Inokawa, and M.~Notomi,
  \enquote{Fast all-optical switching using ion-implanted silicon photonic
  crystal nanocavities,} \apl \textbf{90}, 031115 (2007).

\bibitem{Gibbs:79}
H.~M. Gibbs, T.~N.~C. Venkatesan, S.~L. McCall, A.~Passner, A.~C. Gossard, and
  W.~Wiegmann, \enquote{Optical modulation by optical tuning of a cavity,}
  \apl \textbf{34}, 511--514 (1979).

\bibitem{Tanabe:05}
T.~Tanabe, M.~Notomi, S.~Mitsugi, A.~Shinya, and E.~Kuramochi,
  \enquote{All-optical switches on a silicon chip realized using photonic
  crystal nanocavities,} \apl \textbf{87}, 151112 (2005).

\bibitem{Xu:05}
Q.~Xu, V.~R. Almeida, and M.~Lipson, \enquote{Micrometer-scale all-optical
  wavelength converter on silicon,} \ol \textbf{30}, 2733--2735 (2005).

\bibitem{Fushman:07}
I.~Fushman, E.~Waks, D.~Englund, N.~Stoltz, P.~Petroff, and
  J.~Vu\v{c}kovi\'{c}, \enquote{Ultrafast nonlinear optical tuning of photonic
  crystal cavities,} \apl \textbf{90}, 091118 (2007).

\bibitem{Vahala:03}
K.~J. Vahala, \enquote{Optical microcavities,} \nat \textbf{424}, 839--846
  (2003).

\bibitem{Born:97}
M.~Born and E.~Wolf, \emph{Principles of Optics} (Cambridge University Press,
  1997).

\bibitem{Chin:96}
A.~Chin, K.~Y. Lee, B.~C. Lin, and S.~Horng, \enquote{Picosecond photoresponse
  of carriers in {Si} ion-implanted {Si},} \apl \textbf{69},
  653–655 (1996).

\bibitem{Segschneider:97}
G.~Segschneider, T.~Dekorsy, H.~Kurz, R.~Hey, and K.~Ploog, \enquote{Energy
  resolved ultrafast relaxation dynamics close to the band edge of
  low-temperature grown {GaAs},} \apl \textbf{71}, 2779 (1997).

\bibitem{InfluenceQD}
The maximum unsaturated unbroadened refractive index change of the dots amounts
  to only $10^{-8}$, while the absorption at resonance is less than $50$
  cm$^{-1}$.

\bibitem{EuserWoodpile:08}
T.~G. Euser, A.~J. Molenaar, J.~G. Fleming, B.~Gralak, A.~Polman, and W.~L.
  Vos, \enquote{All-optical octave-broad ultrafast switching of {Si} woodpile
  photonic band gap crystals,} \prb \textbf{77}, 115214 (2008).

\bibitem{EuserThesis:07}
T.~G. Euser, \enquote{Ultrafast optical switching of photonic crystals,} Ph.D.
  thesis, University of Twente, ISBN 978-90-365-2471-1,
  www.photonicbandgaps.com (2007).

\bibitem{MeasurementtauP}
$\tau_P$ denotes the FWHM of the pulse intensity, and was measured in an
  autocorrelator.

\bibitem{Euser:05}
T.~G. Euser and W.~L. Vos, \enquote{Spatial homogeneity of optically switched
  semiconductor photonic crystals and of bulk semiconductors,} J. Appl. Phys.
  \textbf{97}, 043102 (2005).

\bibitem{Blakemore:82}
J.~S. Blakemore, \enquote{Semiconducting and other major properties of gallium
  arsenide,} J. Appl. Phys. \textbf{53}, R123 -- R181 (1982).

\bibitem{Fern:71}
R.~E. Fern and A.~Onton, \enquote{Refractive index of {AlAs},} J. Appl. Phys.
  \textbf{42}, 3499 -- 3500 (1971). Note that the refractive index is given by
  Re($\epsilon^{1/2}$).

\bibitem{Hense:97}
S.~G. Hense and M.~Wegener, \enquote{Ultrafast switch-off of a vertical-cavity
  semiconductor laser,} \prb \textbf{55}, 9255--9258 (1997).

\bibitem{Cho:90}
G.~C. Cho, W.~K\"utt, and H.~Kurz, \enquote{Subpicosecond time-resolved
  coherent-phonon oscillations in {GaAs},} \prl \textbf{65},
  764--766 (1990).

\bibitem{LinearizationN}
The relative error of this linearization is less than $2 \%$ for $N_0 < 3
  \times 10^{25}$ m$^{-3}$, where $N_0$ is the initial carrier density.

\bibitem{ChangeLB}
Independent calculations confirmed that the $L_B$ changes by less than $10 \%$
  for our carrier densities.

\bibitem{Gerard:90}
J.~M. G\'erard, B.~Sermage, L.~Bergomi, and J.~Y. Marzin,
  \enquote{Differentiation of the non radiative recombination properties of the
  two interfaces of {MBE} grown {GaAs-GaAlAs} quantum wells,} Superlattices
  Microstruct. \textbf{8}, 417--419 (1990).

\bibitem{Tanabe:08}
T.~Tanabe, H.~Taniyama, and M.~Notomi, \enquote{Carrier diffusion and
  recombination in photonic crystal nanocavity optical switches,} J. Lightwave
  Tech. \textbf{26}, 1396--1403 (2008).

\bibitem{Notomi:08}
M.~Notomi, E.~Kuramochi, and T.~Tanabe, \enquote{Large-scale arrays of
  ultrahigh-{Q} coupled nanocavities,} Nat. Phot. \textbf{2}, 741--747 (2008).

\bibitem{Sabbah:02}
A.~J. Sabbah and D.~M. Riffe, \enquote{Femtosecond pump-probe reflectivity
  study of silicon carrier dynamics,} \prb \textbf{66}, 165217 (2002).

\end{thebibliography}
\end{document}